\documentstyle[12pt,a4,amstex,epsfig]{article}
\begin{document}

\begin{flushright}
MPI-PhT/96-14(extended version)\\
July 1996\\
\end{flushright}
\begin{center}
\large {\bf A Note on QCD Corrections to
$A^b_{FB}$ using Thrust to determine the $b$-Quark
Direction}\\
\mbox{ }\\
\normalsize
\vskip3cm
{\bf Bodo Lampe}
\vskip0.3cm
Max Planck Institut f\"ur Physik \\
F\"ohringer Ring 6, D-80805 M\"unchen \\
\vspace{3cm}

{\bf Abstract}\\
\end{center}
I discuss one-loop QCD corrections to the
forward backward asymmetry of the $b$-quark in a way appropriate
to the present experimental procedure. I try to give insight
into the structure of the corrections and elucidate some questions
which have been raised by experimental experts. Furthermore,
I complete and comment on results given in the literature.

\newpage

The forward backward asymmetry of the $b$ quark is one of the
most interesting quantities which has been measured at LEP.
It is defined as the ratio
$$
A_{FB} = {\sigma_{F-B}\over\sigma_{F+B}}
$$
and in lowest order is given by $A^{\rm Born}_{FB}=3{v_ba_b\over
v^2_b+a^2_b}{v_ea_e\over v^2_e+a^2_e}$ and therefore sensitive
to the couplings of electron and $b$ quark
to the $Z$. Even more interesting
might be the measurement of the combined left right forward
backward asymmetry $A^{LR}_{FB}={(\sigma_{F-B})_L-(\sigma_{F-B})_R\over
\sigma_{\rm total}}$ projected by SLC because in lowest order
it involves the $b$ quark couplings $v_b$ and $a_b$ only
\cite{R1}.
 
For a precision measurement of these quantities the
understanding of higher order corrections is very important.
Oneloop (electroweak and QCD) corrections have been reviewed in
\cite{R1} and \cite{R2} and twoloop QCD corrections
have been calculated in \cite{R3}.
Usually, these results are presented under the assumption
that the $b$-quark direction can be experimentally precisely
determined. However, with the existing detectors this is not
the case. Instead, the LEP experiments apply the following
procedure \cite{R4}:
\begin{itemize}
\item Events in which the $b$ (or the $\bar b$) decays
semileptonically, $b\to c\mu^-\bar\nu (\bar b\to \bar c\mu^+\nu)$
are selected.
\item For these events the thrust axis $T$ is determined as
the $max_{\vec n}\sum_i
\vert \vec p_i \vec n_i \vert$ where
the sum is over all charged momenta $\vec p_i$
in the event.
\item The orientation of the thrust axis is chosen in
such a way that $\vec T
\vec \mu^-$ is positive
(resp.\
$\vec T\vec\mu^+$ is negative).
Then the event
is
counted forward if $\vec T$ points in the forward
direction $(\vec T\vec e^->0)$
and backward if $\vec T$ points in the backward
direction $(\vec T\vec e^-<0)$.
\end{itemize}
 
This procedure
will be called the
``$\underline{T \hbox{ procedure}}$''
in the following (in contrast to the
``$\underline{b \hbox{ procedure}}$''
where the $b$ quark is used to determine the asymmetry).
The $T$ procedure has several deficiencies:

\noindent
First, due to the missing momentum of the
neutrino, $\vec T$ is not the ``true'' thrust axis.

\noindent
Secondly, due to the nature and kinematics of the $b$
decay, there are events where the $\mu^-$ goes forward while
the $b$ goes backward (and vice versa).

\noindent
Thirdly, gluon emission may spoil the connection between
thrust axis and $b$-quark direction.
 
These deficiencies must all be corrected for. They can be
corrected for separately. In this note we concentrate on point 3.
The corrections to items no.\ 1 and 2 can be made using existing
Monte Carlo programs.
Note that in this procedure
the muon is only used to determine the hemisphere in
which its parent quark is to be expected. One may ask why not
determine the asymmetry of the muon (``$\mu$ procedure'').
The answer is that the muon asymmetry will be notably smaller
than the $b$ asymmetry, because there is a loss of the
original information through the missing neutrino. The $\mu$ procedure
is therefore worse than the $T$ procedure.
 
Before I address item no.\ 3 I will consider the structure of
oneloop QCD corrections to $A_{FB}$ in general. For simplicity,
I will neglect mass terms $O(m_b/m_Z)$ in the following.
In addition to $e^+e^-\to b\bar b$ one has
processes $e^+e^-\to b\bar b g$.
When they are included the QCD effect can be written as an
overall correction factor
$$
A_{FB}=A^{{\rm Born}}_{FB}\cdot (1+c{\alpha_s\over\pi})
$$
which we decompose as
$$
1+c {\alpha_s\over\pi}={1+{\alpha_s\over\pi}(p_2+p_3)\over
                    1+{\alpha_s\over\pi}(q_2+q_3)}
$$
i.e.\ we write it as a correction factor $p$ to $\sigma_{F-B}$
defined by a correction factor $q$ to $\sigma_{F+B}=\sigma_{{\rm total}}$.
Both $p$ and $q$ can be split into a 2 jet and a 3 jet piece in the
sense that one can split $\sigma_{F\pm B}$ in a 2 jet and a 3 jet piece,
$$
\sigma_{F\pm B}=\sigma_{F\pm B}^2(y)+\sigma^3_{F\pm B}(y)\/,
$$
with an invariant
mass cut $y$ to define the jets.
$p_{2,3}(y)$ are defined by
$$
\sigma^2_{F-B}(y)=\sigma_{F-B}^{{\rm Born}}(1+{\alpha_s\over\pi}p_2)\quad 
$$
$$
\sigma^3_{F-B}(y)={\alpha_s\over\pi}p_3\sigma_{F-B}^{{\rm Born}} 
$$
$$
\sigma^2_{F+B}(y)=\sigma_{F+B}^{{\rm Born}}(1+{\alpha_s\over\pi}q_2) 
$$
$$
\sigma^3_{F+B}(y)={\alpha_s\over\pi}q_3\sigma_{F+B}^{{\rm Born}}
$$
Note that the sums $p_2+p_3$ and $q_2+q_3$ are independent of $y$.
One could define a forward backward asymmetry based on 2 jet resp.\
3 jet events only
$$
A^2_{FB}(y)={\sigma^2_{F-B}\over\sigma^2_{F+B}}=A^{{\rm Born}}_{FB}
      (1+{\alpha_s\over\pi}(p_2-q_2)) 
$$
$$
A^3_{FB}(y)={\sigma^3_{F-B}\over\sigma^3_{F+B}}=
      {p_3\over q_3} A^{{\rm Born}}_{FB}
$$
but we shall not consider these quantities in the following.
No simple relation holds between
$A_{FB}$, $A^2_{FB}$ and $A^3_{FB}$.
 
The functions $p_{2,3}(y)$ and $q_{2,3}(y)$ have been given
in the literature \cite{R3}
and I do not want to repeat them here because
I am only interested in the inclusive correction factor $c$.
Within the $b$ procedure one has $c=-1$ (for $m_b=0$) and
$c\approx -0.8$ (for $m_b=4.7$ GeV). It is a question of some
interest to know the value of $c$ for the $T$ procedure,
too. To determine this value we shall work on the parton
level and mimic the $T$ procedure on the parton level.
On the parton level the role of the muon direction is
played by the $b$ quark direction and the thrust dirction
$\vec t$ is given by the parton with the maximum
energy. In lowest order and in the exact 2 jet limit $(y\to 0)$
the thrust direction and the $b$ quark direction are identical
so that no correction needs to be applied (as compared to the $b$
procedure). A difference arises, however, in the 3 jet region,
where $\vec t$ can be either
$\vec b$, $\bar b$
or
$\vec g$.
In $O(\alpha_s)$ the $T$ and $b$ procedure are equivalent
only
in the strict 2 jet limit $y\to 0$. An event is forward if either
$\vec t\vec b >0$ and
$\vec t\vec e^- >0$ or
$\vec t\vec  b<0$ and
$\vec t\vec e^-<0$,
and backward otherwise. We
have
used this procedure and applied it to the QCD matrix element for
the process $e^+e^-\to b\bar b g$. One obtains the following results:
$$
c(T\hbox{ procedure, }m_b=0)=-0.893
$$
This number can be decomposed into a 2 jet and 3 jet contribution.
$$
c=(k_2+k_3){C_F\over 2}\qquad\hbox{with}\qquad
{C_F\over 2}k_{2,3}=(p-q)_{2,3}
$$
The colour factor $C_F=4/3$ has been introduced for convenience.
The results for $k_3$ are given in the table, both for the $b$ and the
$T$ procedure for several values of $y$, assuming $m_b=0$.
$k_2={3\over 2}c-k_3$ vanishes
for $y\to 0$ because in the $m_b=0$ theory there is no contribution
from the virtual gluon exchange diagram $e^+e^-\to b\bar b$.
Of course it is desirable to have the
$0(m_b)$ dependence of $c$ in the $T$ procedure.  
This is done in a forthcoming publication.
 
In summary one may state: when going from 
the $b$ procedure ($c=-1$) to the $T$
procedure $(c=-0.893)$ one gets a correction of
about 10\% to the correction, i.e.\ the effect is small and
irrelevant on the basis of the present experimental accuracy
and only important for some future precision experiment. 
This statement remains true if $0(m_b)$ corrections are included.

\begin{table}
\begin{tabular}{l l l}
y       &$k_3$ (b procedure)   &$k_3$(T procedure) \\
\hline
0       &-1.500                &-1.340 \\
0.001   &-1.474                &-1.328 \\
0.005   &-1.393   &-1.283 \\
0.01    &-1.313   &-1.225 \\
0.02    &-1.183   &-1.148 \\
0.04    &-0.974   &-0.996 \\
0.06    &-0.811   &-0.875 \\
0.08    &-0.671   &-0.753 \\
0.10    &-0.553   &-0.653 \\
0.12    &-0.451   &-0.557 \\
0.14    &-0.363   &-0.472 \\
0.16    &-0.288   &-0.388 \\
0.20    &-0.165   &-0.244 \\
1       &0        &0 \\
\end{tabular}
\end{table}
\bigskip  

\vskip1cm
\noindent {\bf Acknowledgements:} This work was done while
I was visiting CERN. I would like to thank Guido Altarelli for
his encouragement and Roberto Tenchini and Duccio Abbaneo
for several discussions.
 
\newpage

{\bf Appendix:Some Details of the Calculation}
 
We start with 3 different representations of the cross section
$$
{d\sigma\over d\cos\theta_b}=
\sigma^b_V \cos\theta_b+{3\over 4}\sigma^b_L\sin^2\theta_b+{3\over 8}\sigma^b_U
(1+\cos^2\theta_b)
$$
+ aximuthal terms  ($\phi_b$);
$$
{d\sigma\over d\cos\theta_{\bar b}}=
\sigma^{\bar b}_V \cos\theta_{\bar b}+{3\over 4}\sigma^{\bar
b}_L\sin^2\theta_{\bar b}
+{3\over 8}\sigma^{\bar b}_U
(1+\cos^2\theta_{\bar b})
$$
+ aximuthal terms  ($\phi_{\bar b}$);
$$
{d\sigma\over d\cos\theta_g}=
\sigma^g_V \cos\theta_g+{3\over 4}\sigma^g_L\sin^2\theta_g+{3\over 8}\sigma^g_U
(1+\cos^2\theta_g)
$$
+ aximuthal terms  ($\phi_{g}$);
where
$$
\sigma^i_V={3\over 8} J\sigma_o{\alpha_s\over 2\pi}C_FB^i_V
$$
$$
\sigma^i_{U+L}=R\sigma_o{\alpha_s\over 2\pi} C_F B_{UL}
$$
and
$$
B_{UL}={x^2_1 +x^2_2\over y_{13}y_{23}}
$$
$$
B^i_V={x^2_1\cos\theta_{1i}-x^2_2\cos\theta_{2i}\over y_{13}y_{23}}
$$
$\cos\theta_{ij}=1$ for i=j and 
$$
\cos\theta_{ij}=1+{2\over x_ix_j}-{2\over x_i}-{2\over x_j} 
$$
for $i \neq j$.
The $x_i$ are the normalized energies of partons $i$,
$x_1+x_2+x_3=2$, $y_{13}=1-x_2$ etc. as usual.
$\sigma_o$ is the tree level cross section for $e^+e^-\rightarrow 
\mu^+\mu^-$ (photon exchange only).
Note that at leading order
$e^+e^-\to b\bar b$
$$ A^0_{FB}={3 J\over 8R}\/.
$$
Expressions for J and R can be found, for instance, in 
Nachtmann's book. 
In \cite{R3} it was shown that in the $b$ procedure the QCD
correction factor can be written as
$$
1+{\alpha_s\over 2\pi}C_F(B^b_V-B_{UL})_{{\rm all}}=
1-{3\over 2}{\alpha_s\over 2\pi}C_F
$$
where $(~~)_{{\rm all}}$ denotes
$\int^1_0 dx_1\int^1_0 dx_2\theta (x_1+x_2-1)$
and the singularities of $B^b_V$ and $B_{UL}$ for
$y_{13}\to 0, y_{23}\to 0$ drop out in the difference $B^b_V-B_{UL}$.
 
In the $T$ procedure the QCD correction factor is given by
$$
1+{\alpha_s\over 2\pi}C_F\Bigl\{(B^b_V -B_{UL})_{x_1>}+
(-B^{\bar b}_V-B_{UL})_{x_2>}+(-B^g_V-B_{UL})_{x_3>}\Bigr\}
$$
where
$(~~~)_{x_1>}$ denotes
$\int\limits^1_0dx_1\int\limits^1_0dx_2\theta (x_1+x_2-1)
\theta(x_1-x_2)\theta(x_1-x_3)$ etc.
 
Numerically one finds
$$
(B^b_V-B_{UL})_{x_1>} =(-B^{\bar b}_V-B_{UL})_{x_2>}=-0.21
$$
$$
(-B^g_V-B_{UL})_{x_3>}=-B_{UL}\Big|_{x_3>}=-0.92
$$
For comparison we also give here
$$
(B^b_V-B_{UL})_{x_2>}=-0.74
$$
$$
(B^b_V-B_{UL})_{x_3>}=-0.55
$$
The difference between $b$ and $T$ procedure is given by
$$
{\alpha_s\over 2\pi}C_F [ (B^b_V+B^{\bar b}_V)_{x_2>} +
B^b_V\Big|_{x_3>} ]=
{\alpha_s\over 2\pi}C_F(-0.5273 +0.3675)
$$
$$
=(-0.160\pm 0.003) {\alpha_s\over 2\pi}C_F
$$

\end{document}